\documentclass[twocolumn,english]{article}
\usepackage[T1]{fontenc}
\usepackage[latin9]{inputenc}
\usepackage{geometry}
\usepackage{textcomp}
\usepackage{graphicx}
\usepackage{babel}
\begin{document}
\date{}

\twocolumn[
\begin{@twocolumnfalse} 

\title{\textbf{Light Ion Accelerating Line (L3IA): Test Experiment at ILIL-PW }}

\maketitle
\author {L.A.Gizzi\textsuperscript{a,b}, }
\author {F. Baffigi\textsuperscript{a},}
\author {F. Brandi\textsuperscript{a},}
\author {G. Bussolino\textsuperscript{a,b},}
\author {G. Cristoforetti\textsuperscript{a},}
\author {A. Fazzi \textsuperscript{c},}
\author {L. Fulgentini\textsuperscript{a},}
\author {D. Giove\textsuperscript{d},}
\author {P. Koester\textsuperscript{a},}
\author {L. Labate\textsuperscript{a,b},}
\author {G. Maero\textsuperscript{e},}
\author {D. Palla\textsuperscript{a},}
\author {M. Romé\textsuperscript{e},}
\author {P. Tomassini\textsuperscript{a}.}

$\,$

\hphantom{A}\textsuperscript{a}\textit{ Intense Laser Irradiation Laboratory, INO-CNR, Pisa, Italy.}

\hphantom{A}\textsuperscript{b}\textit{ INFN, Sez. Pisa, Italy.}

\hphantom{A}\textsuperscript{c}\textit{ Dipartimento di Energia, Politecnico di Milano and INFN, Sezione di Milano, Italy.}

\hphantom{A}\textsuperscript{d}\textit{ INFN-LASA, Segrate, Italy}

\hphantom{A}\textsuperscript{e}\textit{Università di Milano and INFN Sez. Milano, Italy.}

$\,$

\begin{abstract}
The construction of a novel Laser driven Light Ions Acceleration Line
(L3IA) is progressing rapidly towards the operation, following the
recent upgrade of the ILIL-PW laser facility. The Line was designed
following the pilot experimental activity carried out earlier at the
same facility to define design parameters and to identify main components
including target control and diagnostic equipment, also in combination
with the numerical simulations for the optimization of laser and target
parameters. A preliminary set of data was acquired following the successful
commissioning of the laser system >100 TW upgrade. Data include output
from a range of different ion detectors and optical diagnostics installed
for qualification of the laser-target interaction. An overview of
the results is given along with a description of the relevant upgraded
laser facility and features.

$\,$

$\,$
\end{abstract}
\end{@twocolumnfalse}
] 

\section{Introduction}

Novel acceleration techniques based on ultra intense lasers are evolving
rapidly from scientific exploration to applications, relying on established
and extensively investigated \cite{Daido} acceleration processes
like the Target Normal Sheath Acceleration (TNSA)\cite{Snavely}. 

Examples of applications include injectors for high power ion beams,
neutron generation\cite{roth}, probes for fast evolving phenomena
like the ultrafast charging of laser-heated samples\cite{ahmed} and
space radiation studies and electronic components testing\cite{hidding}.
Applications with potential impact on industry and cultural heritage
like the proton-induced X-ray emission spectroscopy (PIXE) may be
applicable with currently achievable TNSA performances and may strongly
benefit from the compactness of a multi-MeV laser-driven ion source\cite{barberio}.

Laser-based applications requiring multi MeV ions are being developed
for industrial use, in view of the ongoing evolution of the next generation
of Joule-scale laser drivers in the sub-100 fs domain, which may become
attractive\cite{Noaman} for their higher repetition rate, potentially
reaching the 100 Hz or even the kHz range. In fact, with the ongoing
transition to an extensive use of diode-pumping in high power lasers\cite{mason},
high repetition rate and higher efficiency TNSA drivers may soon be
available enabling laser-driven high average power sources to become
finally commercially available. 
\begin{figure}
\includegraphics[scale=0.18]{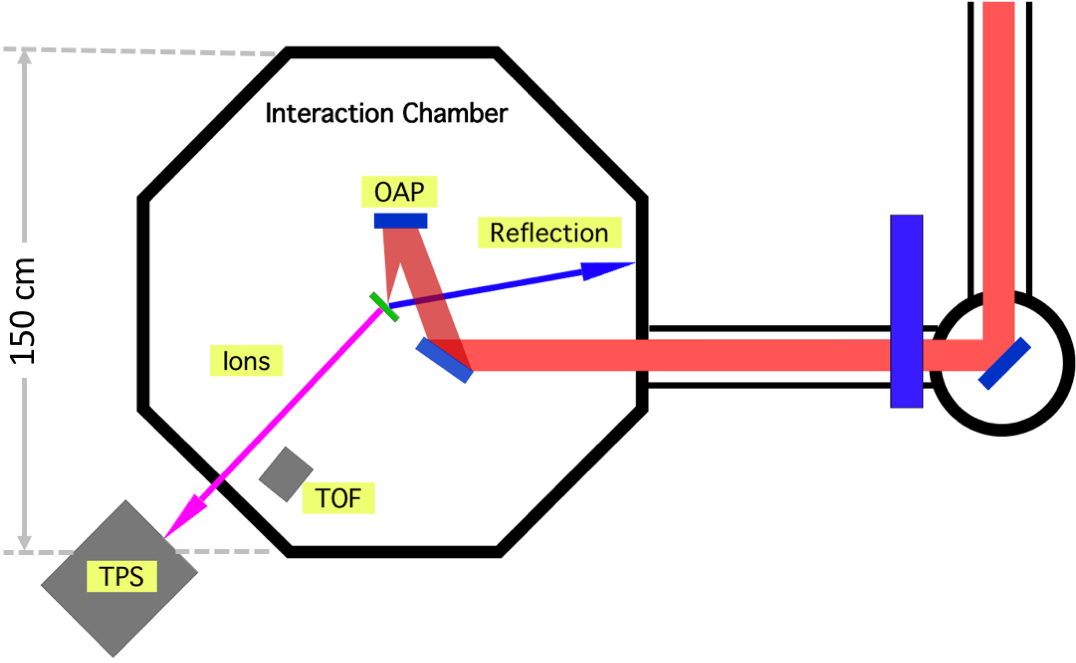}\caption{Schematic experimental setup showing the octagonal target chamber,
the off-axis parabola and the main diagnostics.\label{fig:setupcamera}}
\end{figure}

At the same time, great attention is being dedicated to the control
of acceleration parameters to enhance TNSA performances, including
energy cut-off, beam divergence, charge, emittance. Target optimisation
and engineering, looking at different properties of surface, geometry
and conductivity are becoming crucial in this effort, with nano-structured
targets emerging as a potential breakthrough in table-top laser-driven
ion sources development\cite{blanco}. Finally, post acceleration
is being tackled with special attention to selection, collimation\cite{romano}
and injection in secondary acceleration structures, even using miniature
target-driven guiding devices\cite{kar}.

Here we describe the preliminary results of the commissioning experiment
of a new Line for Laser-Driven Light Ions Acceleration (L3IA) with
the purpose of establishing an outstanding beam-line operation of
a laser-plasma source in Italy taking advantage of the results achieved
so far in this field by the the precursor activity\cite{agosteo}
and based upon experimental campaigns and numerical modeling. The
beam-line will operate in the parameter range of ion acceleration
currently being explored by leading European laboratories in this
field and will provide an advanced test facility for the development of
laser driven ion sources. The project includes a complete set of work-packages,
including the interface with the ILIL-PW facility, the beam line scheme
comprising targets, laser beam focusing, and diagnostic devices dedicated
to both the laser-plasma interaction and the ion beam detection and
characterisation. Numerical modelling is also included, with the specific
tasks of providing basic predictive simulations for the baseline parameters
of the beam line and also allowing investigation of advanced target
and laser configurations. Provision is also made for specific application
cases, including radiobiological testing and cultural heritage applications.

\section{Experimental set up}

The experiment was carried out at the Intense Laser Irradiation Laboratory
using the ILIL-PW Ti:Sa laser and interaction facility using laser
pulse parameters related to the phase 1 configuration described in
Ref.\cite{gizzi AP}. Preliminary results obtained using the laser
pulse at the output of the front-end can be found in Ref.\cite{gizzi AP}.
In the same reference, an overview of the ILIL-PW facility and a summary
of the main laser parameters are also presented. In the experiment presented here
the pulse duration was 30 fs and the pulse energy was 3 J on target.

A schematic view of the experimental set up is given in Fig.(\ref{fig:setupcamera}).
The 100 mm diameter beam was focused by an F/4.5 Off-Axis Parabolic (OAP) mirror 
with an angle of incidence of 15\textdegree{}. The focal spot was elliptical, with an average 
diameter of 4.4 $\mathrm{\mu m}$ (FWHM) and an intensity in excess 
of $\mathrm{1.6\times10^{20}}\:\mathrm{W/cm^{2}}$ ($\mathrm{\mathrm{a_{0}=8.6}}$).
The target was mounted in a remotely controlled motorized support with
a sub-micrometer resolution, capable of XYZ translation and azimuthal
rotation around the vertical axis. 

\begin{figure}
\includegraphics[scale=0.25]{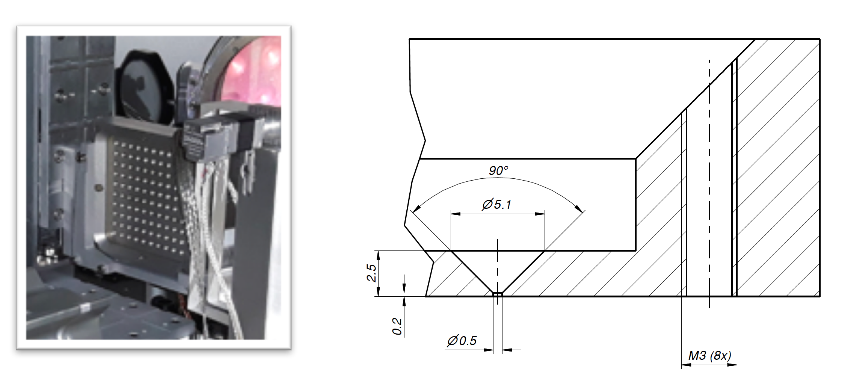}\caption{(left) Target mount showing the 500 $\mathrm{\mu u}$m holes array.
(right) Detail showing the conical hole geometry. \label{fig:montaggio}}
\end{figure}

As shown in Fig.(\ref{fig:montaggio}), the target mount consisted
of a solid steel frame machined to leave access to the surface of
the foil from both sides through a set of 500 $\mu$m diameter holes
with conical aperture to allow oblique laser incidence on target.
The whole mount was designed to enable a 100-mm range of positioning
and withstand a load of up to 500 N in all directions of motion. These
specifications ensure that the scanning of targets up to 100 mm $\times$
100 mm can be accomplished, enabling a large number of laser shots
to be fired on a given target before target replacement is required.
In the measurements discussed here the target consist of a $\mathrm{10}$
$\mathrm{\mu m}$ Al foil.

\begin{figure}
\includegraphics[scale=0.6]{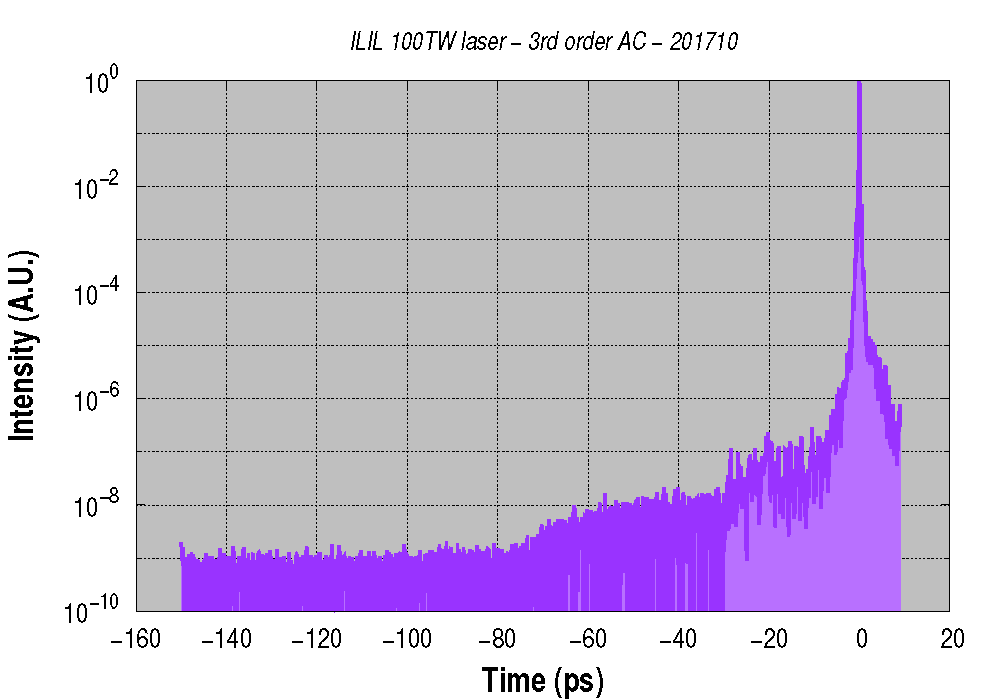}\caption{Cross-correlation curve of the laser pulse. \label{fig:Cross-correlation-curve-of}}

\end{figure}

Special attention was dedicated, during the experiment, to establish
target integrity at the time of arrival of the main pulse on target
which strongly depends on the temporal profile of the laser pulse\cite{baffigi}.
A cross-correlation curve of the laser pulse taken with the Sequoia
(Amplitude Technologies) is shown in Fig.(\ref{fig:Cross-correlation-curve-of}).
According to this plot, the laser contrast is greater than $\mathrm{10^{7}}$
up to 10 ps before the peak of the pulse.
 A detailed modelling of laser-target interaction with such a laser
temporal profile is in progress, but we can anticipate that with the
measured laser contrast, no major pre-plasma formation occurs, ensuring
bulk target survival at the peak of the pulse.

These circumstances were further confirmed by optical spectroscopy
of the light scattered in the specular direction, collected shot by
shot using a F/5 collecting lens placed in vacuum to collimate scattered light 
outside the target chamber. Light was than attenuated using neutral filters and 
rejecting filters at 800 nm and focused on the tip of a fiber coupled to a spectrometer.
This set up enabled detection of second harmonic emission, $\mathrm{\mathrm{2\omega_{L}}}$, 
and $\mathrm{\mathrm{\left(3/2\right)\omega_{L}}}$ of the incident
laser light scattered in the specular direction. Such components are associated to the coupling of the
laser light at the critical density and at the quarter critical density
respectively\cite{gizzi xray}. 

In fact, the formation of even a very small pre-plasma before of the
arrival of the main pulse can provide suitable conditions for the
growth of stimulated instabilities including the Stimulated Raman
Scattering and the Two Plasmon Decay (see \cite{barberio} and references
therein). Electron plasma waves at $\mathrm{\mathrm{\omega_{L}/2}}$
generated by the instabilities can couple non-linearly with incident
laser light and give rise to $\mathrm{\mathrm{\left(3/2\right)\omega_{L}}}$
emission. This emission is therefore a signature of the presence of
even a small pre-plasma.

Second harmonic emission in the specular direction is generated by
the non-linear interaction of the main laser pulse at the critical
density\cite{gizzi xray 2}. Therefore, second harmonic emission can
be taken as a signature of the presence of a critical density layer
in the plasma at the time of interaction of the main pulse, a prerequisite
for the interaction with an over-dense target and the occurrence of
TNSA. 

\begin{figure}
\includegraphics[scale=0.35]{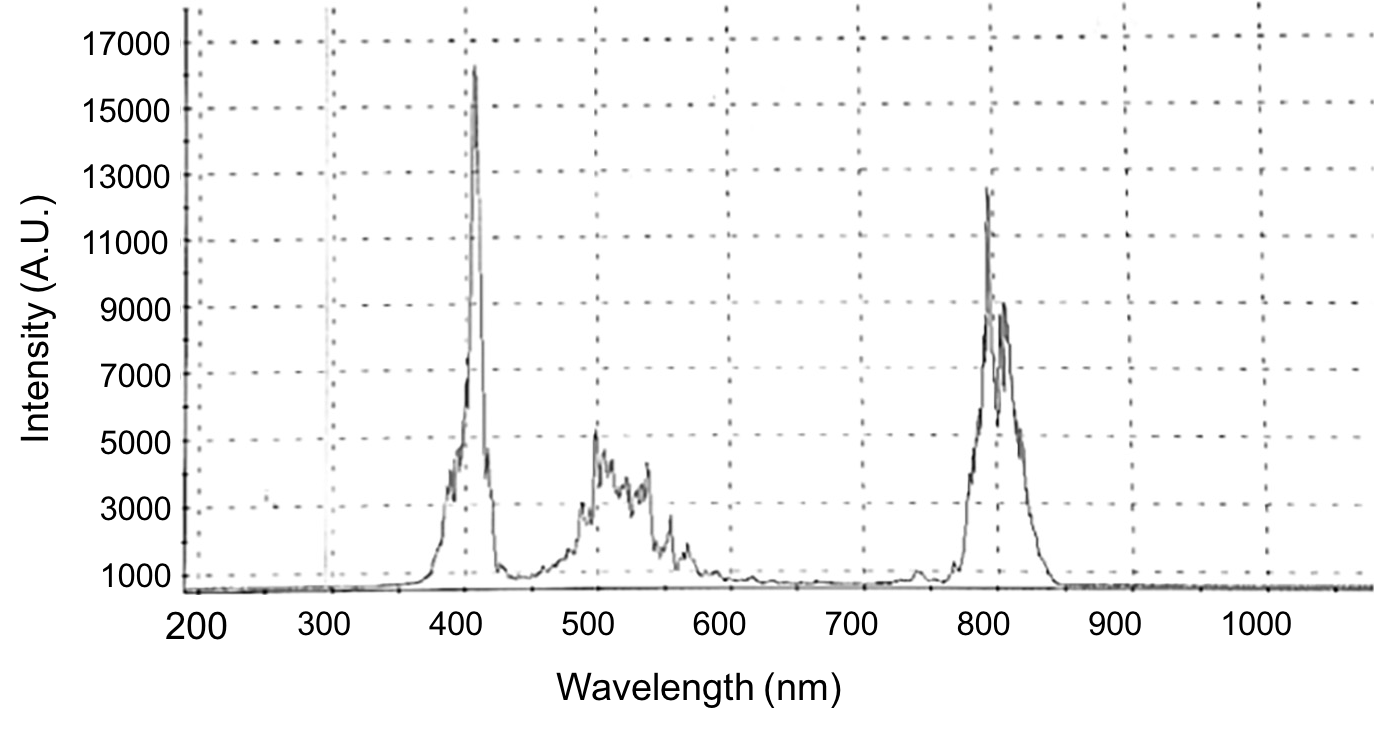}
\caption{A typical spectrum obtained with the target placed near the best focus.
The second harmonic component at 400 nm is used to monitor the survival
of the target at the peak of the pulse. \label{fig:A-typical-spectrum}}
\end{figure}

As shown by Fig.(\ref{fig:A-typical-spectrum}), in our experiments,
in spite of the increase of the $\mathrm{\mathrm{\left(3/2\right)\omega_{L}}}$
intensity compared to previous experimental campaign at 10 TW Ref.\cite{gizzi AP},
the intensity of the $\mathrm{\mathrm{2\omega_{L}}}$ emission remains
always significant, indicating that the laser contrast in the best
focus is sufficient to ensure survival of an overdense target.

\section{Results and discussion}
A range of diagnostics were used in our experiments to measure ion
acceleration, including radio-chromic films (GAF), CR39, Thomson Parabola,
and Time of Flight (TOF) diamond detectors. Thomson Parabola Spectrometer
(TPS) and a TOF detector were used simultaneously so that a cross-comparison
of the signals obtained from the two devices was possible. This was
done in view of a possible use of the diamond detector for on-line
direct detection of accelerated ions during normal operation. A detailed
discussion of all these measurements with different detectors is given
elsewhere\cite{Gizzi1,altana}. Here, we focus our attention on the
presentation of the preliminary results of TOF and TPS signals obtained
during the currently operating L3IA phase 1 configuration. 

\begin{figure}
\includegraphics[scale=0.27]{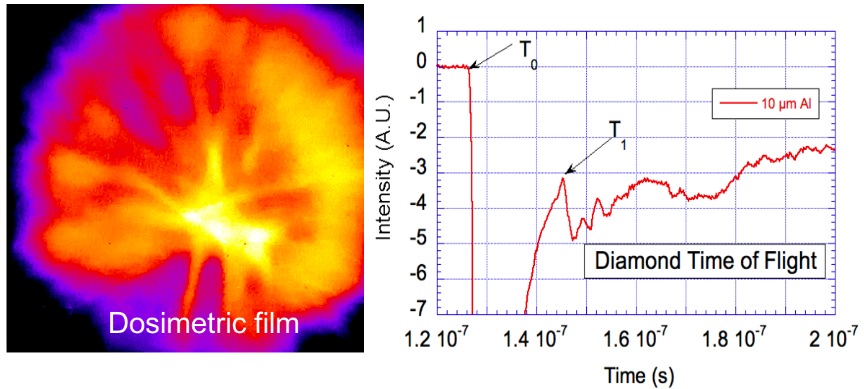}
\caption{(left) The signal from the GAFchromic$\mathrm{^{TM}}$ EBT-3 film
placed at 100 mm from the target rear surface. (right) Signal of the
diamond Time of Flight (TOF) showing the ion signal from the irradiation
of a 10-$\mathrm{\mu}$m thick aluminium target. T=0 is arbitrary
in this plot. The strong emission in the range $\mathrm{\mathrm{t=[T_{0}-T_{1}]}}$
is the signal due to fast electrons. For $\mathrm{\mathrm{t>T_{1}}}$
the signal is due to ions detection.\label{fig:The-signal-from}}
\end{figure}

A typical GAF image obtained with a 10 $\mathrm{\mu}$m thick Al target
is shown in Fig.(\ref{fig:The-signal-from})(left), showing an intense
on-axis spot, surrounded by a broader signal.

Fig.(\ref{fig:The-signal-from})(right) shows the plot of the TOF
signal obtained with the diamond detector from the irradiation of
the same 10 $\mathrm{\mu m}$ thick Al target. The TOF detector was
placed at a distance of 40 cm from the target rear side and was filtered
using a 12 $\mathrm{\mu m}$ thick Al foil. The strong peak between
$\mathrm{\mathrm{t=T_{0}}}$ and $\mathrm{\mathrm{t=T_{1}}}$ is attributed
to a combination of X-rays and fast electrons reaching the detector
soon after the interaction\cite{gizzi AP,Gizzi1}. This peak is than
followed by the actual ion signal that starts at $\mathrm{\mathrm{t=T_{1}}}$.
Taking into account the TOF distance and assuming a signal predominantly
due to protons that have the highest charge-to-mass ratio, calculations
yield a high energy cut-off of approximately 6 MeV. 

\begin{figure}
\includegraphics[scale=0.25]{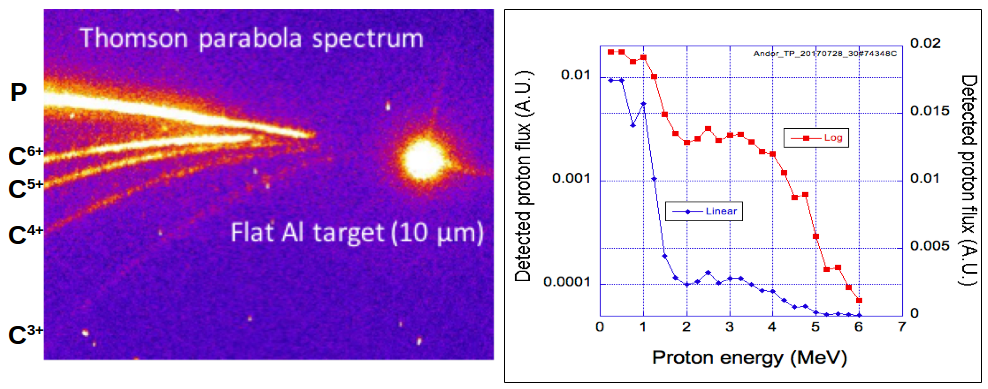}\caption{Image of the Thomson Parabola Spectrometer (TPS) spectrogram from
a 10-$\mathrm{\mu m}$ thick aluminum target showing protons and carbon
ions with different charge states. The proton spectrum is also reported
(right) in logarithmic scale showing the proton cut-off energy of
6 MeV.\label{fig:thomson}}
\end{figure}

For the same aluminium shot, the row TPS spectrogram is also presented
in Fig.(\ref{fig:thomson}), showing the parabolic traces of protons
and carbons with different ionization states. As we can see, the assumption
that the TOF ion signal is predominantly due to protons is confirmed
by TPS results that show a bright. The analysis of the TPS proton
spectrum is presented in Fig.(\ref{fig:thomson})(right). The measured
cutoff proton energy of 6 MeV is in a perfect agreement with the TOF
estimation, also validating the use of our TOF detector for a reliable,
online shot to shot proton energy evaluation. 

\begin{figure}
\begin{raggedright}
\includegraphics[scale=0.17]{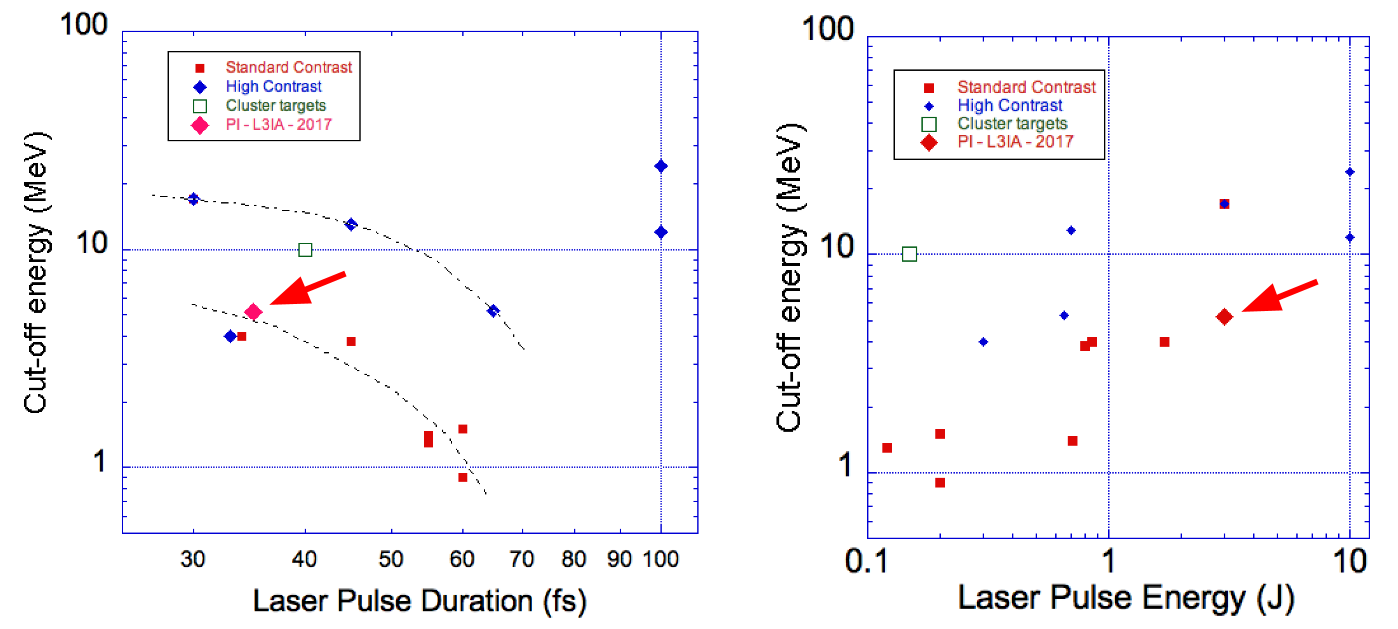}\caption{Summary of data from published TNSA experiments (after\cite{Daido})
showing measured dependence upon laser pulse duration (left) and laser
pulse energy (right). The arrows indicate the cutoff energy of our
experiment.\label{fig:Summary-of-dataconfronto}}
\par\end{raggedright}
\end{figure}

It is interesting to compare the measured cutoff energy with published
results obtained in similar interaction conditions. A summary of published
results relevant for our experimental conditions, taken from\cite{Daido},
is displayed in Fig.(\ref{fig:Summary-of-dataconfronto}) as a function
of laser pulse duration (left) and laser pulse energy (right) showing
the cutoff of our experiment for comparison. These plots show that
the cut-off energy measured in our experiment is in agreement or even
exceeds the cutoff values measured in similar experiments with standard
contrast (no plasma mirror). 

Our preliminary results from this commissioning experiment meet the
expectations from first phase of L3IA as anticipated\cite{gizzi AP},
enabling foreseen applications. Further increase of ion energy and
flux will require fine tuning of laser pulse parameters and target
optimisation.

\section{Conclusions}
In summary, we described the preliminary data obtained during an experiment
dedicated to the commissioning of the new laser-driven light ions
acceleration line (L3IA). Our experiment shows proton acceleration
cut-off energies up to 6 MeV which are in a perfect agreement with
intensity scaling established by previous measurements in similar
interaction conditions. Our data demonstrate overall laser and target
performances of our setup in line with project specification planned
for Phase 1.

\section{Acknowledgements}
This project has received funding from the CNR funded Italian research
Network ELI-Italy and from the L3IA INFN Experiment of CSN5. We gratefully
acknowledge support from the Central Laser Facility for in kind contribution
to the experimental set up described in this experiment.

\end{document}